\documentstyle[12pt]{article}
\begin{document}

\begin{flushright}{CMU-HEP95-05\\
DOE-ER/40682-95}
\end{flushright}

\vspace{.15in}

\begin{center}{{\bf The LSND Experiment and the Zee Model}\\
Lincoln Wolfenstein\\
Department of Physics\\
Carnegie Mellon University\\
Pittsburgh, PA~~15213~~USA}
\end{center}

\vspace{.25in}

A recent experiment LSND at Los Alamos has provided\cite{1} an indication of
$\overline{\nu}_{\mu} - \overline{\nu}_e$ oscillations with $\Delta m^2$ of
order 1ev$^2$ or greater and $\sin^2 2 \theta_{e \mu}$ between $2 \times
10^{-3}$ and
$10^{2}$.  Such a value of $\Delta m^2$ for $\nu_{\mu} - \nu_e$ oscillations
was not expected in the standard see-saw model\cite{2} suggested by SO(10) with
a large mass hierarchy because it leads to too large a value of $m(\nu_{\tau})$
to fit cosmological constraints.  Furthermore within that model the LSND result
is inconsistent with the indications of oscillations from both atmospheric
neutrinos\cite{3} and solar neutrinos.\cite{4}

Here we reconsider an alternative model proposed by Zee.\cite{5}  This
represents the most direct way to produce Majorana neutrino masses in a theory
such as SU(5) in which right-handed neutrinos are totally absent.  The general
features of masses and mixings in this model have been analyzed.\cite{6}  The
mass matrix has the form

$$m_0^2 \left(\begin{array}{lll}
0 & \sigma & \cos  \alpha\\
\sigma & 0 & \sin \alpha\\
\cos \alpha & \sin \alpha & 0\end{array}\right)$$

\noindent where $\sigma$ is of order $(m^2_{\mu}/m_{\tau}^2) \approx 10^{-2}$
to
$10^{-3}$.  The eigenstates to a good approximation are

$$\begin{array}{llllllllllll}
N_1 & = & \nu_e \sin \alpha &  - & \nu_{\mu} \cos \alpha\\
\\
\sqrt{2} N_2 & = & \nu_{\tau} & + & \nu'\\
\\
\sqrt{2} N_3 & = & \nu_{\tau} & - & \nu'\\
\\
\nu' & = & \nu_e \cos \alpha & + & \nu_{\mu} \sin \alpha
\end{array}$$

\noindent The state $N_1$ has a small mass of order $\sigma m^2_0$ whereas
$N_2$ and $N_3$ have masses of order $m^2_0$ with a mass splitting

$$\Delta m^2_{23} = 2 m^2_0 \sigma \sin 2 \alpha. \eqno{(1)}$$

The following features follow from these equations:

1.  There exist two very different values for $\Delta m^2$.  The large value
$\Delta_L  \approx m^2_0$, corresponding to short wave-length oscillations
applies only to $\nu_{\mu} - \nu_e$ oscillations.

2.  There are oscillations of $\nu_{\mu}$ to $\nu_{\tau}$ and $\nu_e$ to
$\nu_{\tau}$ corresponding to a much smaller value of $\Delta m^2$, $\Delta_s$,
given by Eq. (1).

3.  If the mixing angle for the short-wave length $\nu_{\mu} - \nu_e$
oscillations is $\alpha$, then the amplitudes of the oscillations involving
$\Delta_s$ are given by $\cos^2 \alpha$ and $\sin^2 \alpha$, one corresponding
to $
\nu_{\mu} - \nu_{\tau}$ and the other to $\nu_e - \nu_{\tau}$.

4.  There exist two massive neutrinos almost degenerate in mass with masses
given by $\sqrt{\Delta_L}$.

To apply this theory to the LSND experiment we take as an example $\Delta_L$ =
6ev$^2$.  The theory then gives two massive neutrinos with masses each about
2.5 ev; such a scenario has been suggested as being very useful for
cosmology.\cite{7}  Indeed the interpretation of the LSND experiment in terms
of two almost degenerate massive neutrinos has been suggested in various
papers\cite{8}; it is required in the Zee model.

{}From the LSND value of $\alpha$ it follows that either $\nu_e$ or
$\nu_{\mu}$ has almost complete mixing with $\nu_{\tau}$ while the other has an
amplitude of oscillation of order $10^{-3}$.  For our example from Eq. (1),
$\Delta_s$ is of order of magnitude $10^{-2}$ to $10^{-3}$ ev$^2$.  Thus one of
the two possibilities corresponds to complete $\nu_{\mu} - \nu_{\tau}$ mixing
with a value of $\Delta_s$ appropriate to explain the atmospheric neutrino
results.  There is in this case no explanation of the solar neutrino results.
The other possibility provides no explanation of the atmospheric neutrino
results and indeed $\Delta_s$ should be down to $10^{-3}$ ev$^2$ so as not to
exacerate this problem; it does provide a prediction that the solar
neutrino $\nu_e$ flux is reduced by a factor of 2.

While this does not provide a perfect explanation of the solar neutrino
results,
it does explain the gallium and Kamiokande results within their $1 \sigma$
experimental errors.  However, even taking into account the uncertainty in the
$^8B$ flux there is at least a $3 \sigma$ discrepancy with the Davis result.

In conclusion the LSND result combined with the Zee model leads to the
interesting {\it predictions} there are two neutrinos almost degenerate with
masses of interest for cosmology and that a large neutrino oscillation signal
should be seen in either the atmospheric neutrinos or the solar neutrinos.

This research was supported by the U.S. Department of Energy Contract No.
DE-FG02-91ER40682.

\end{document}